\documentclass[convention,e-brief]{aesconf}

\graphicspath{{./}{figures/}}

\usepackage[utf8]{inputenc}
\usepackage{amsmath,cite,url}
\usepackage{graphicx}
\usepackage{color}
\usepackage{booktabs}
\usepackage{multirow}
\usepackage[table]{xcolor}
\usepackage{hyperref}

\titlespacing{\section}{0pt}{0.4em}{0.4em}
\titlespacing{\subsection}{0pt}{0.3em}{0.3em}

\usepackage[numbers,square]{natbib}

\title{WaveBeat: End-to-end beat and downbeat tracking in the time domain} 

\author[]{Christian J. Steinmetz}
\author[]{Joshua D. Reiss}

\affil[]{Centre for Digital Music, Queen Mary University of London}

\correspondence{Christian J. Steinmetz}{c.j.steinmetz@qmul.ac.uk}

\lastnames{Steinmetz and Reiss}

\shorttitle{WaveBeat}

\ebriefnumber{11}

\begin{document}
\twocolumn[
\vspace{-1.2cm}
\maketitle 

\begin{onecolabstract}
Deep learning approaches for beat and downbeat tracking have brought advancements. However, these approaches continue to rely on hand-crafted, subsampled spectral features as input, restricting the information available to the model.
In this work, we propose WaveBeat, an end-to-end approach for joint beat and downbeat tracking operating directly on waveforms.
This method forgoes engineered spectral features, and instead, produces beat and downbeat predictions directly from the waveform, the first of its kind for this task.
Our model utilizes temporal convolutional networks (TCNs) operating on waveforms that achieve a very large receptive field ($\geq$ 30\,s) at audio sample rates in a memory efficient manner by employing rapidly growing dilation factors with fewer layers.
With a straightforward data augmentation strategy, our method outperforms previous state-of-the-art methods on some datasets, while producing comparable results on others, demonstrating the potential for time domain approaches. 
\vspace{0.4cm}
\end{onecolabstract}
]

\section{Introduction}\label{sec:introduction}

Beat tracking involves estimating a sequence of time instants that reflect how a human listener may tap along with a musical piece. 
Downbeat tracking extends this by requiring the estimation not only of the beat locations, but specifically the locations of beats corresponding to the first beat within each bar.
Such a system has applications across music signal processing including automatic transcription \cite{dixon2001automatic}, chord recognition~\cite{di2013automatic}, music similarity~\cite{ellis2007beat}, and remixing~\cite{veire2018raw}.

Early signal processing approaches generally utilized a two-stage pipeline composed of an onset detection function followed by a post-processing phase to determine which onsets correspond to beats, often incorporating musical knowledge~\cite{goto2001audio, laroche2003efficient, dixon2007evaluation, ellis2007beat, davies2007context}. 
In contrast, with the rise of deep learning, systems have adopted predominately data driven approaches. 
Recurrent networks were first shown to be successful in the beat tracking task a decade ago~\cite{bock2011enhanced}, and have now been extended through a number of iterations~\cite{bock2014multi, bock2016joint}.
More recently, convolutional networks have been successful, performing on par with recurrent networks with greater efficiency~\cite{matthewdavies2019temporal}.
Other works have focused on improving performance through the design of domain-inspired features~\cite{durand2015downbeat}, multi-task learning~\cite{bock2019multi, bock2020deconstruct}, or specialized architectures~\cite{di2021downbeat}. 

While these deep learning approaches have demonstrated superior performance, they continue to employ aspects of traditional techniques, namely the use of hand-crafted magnitude spectrograms as input, along with specialized post-processing. 
The use of these features facilitates the construction of models that consider a large context efficiently, but discards a significant amount of information from the time domain signal in the process. 
This discarded information may be relevant for the beat tracking task, but training models directly on audio waveforms likely requires larger models, with more compute and training data~\cite{pons2017end}. 

\begin{figure*}[ht]
    \centering
    \vspace{-0.5cm}
    \includegraphics[width=0.8\linewidth]{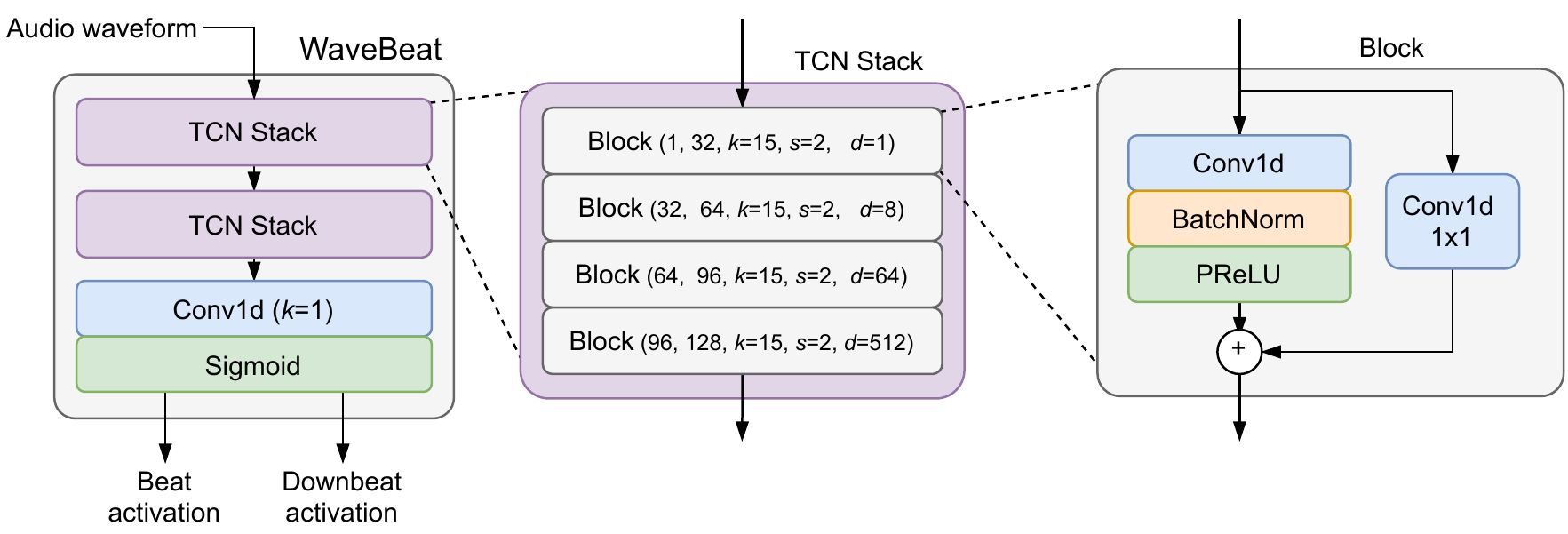}
    \vspace{-0.1cm}
    \caption{The WaveBeat architecture composed of strided 1D convolutions with increasing dilation factors.}
    \vspace{-0.3cm}
    \label{fig:arch}
\end{figure*}

In this work, we investigate learning to jointly predict beat and downbeat events directly from raw audio.
This enables us to forgo engineered features, and take advantage of information within the phase of the input signal, which has been shown to be useful in traditional approaches~\cite{eck2007beat}. 
We also investigate if commonly employed post-processing techniques are actually required, enabling a complete end-to-end system.

We employ a specialized temporal convolutional network (TCN)~\cite{bai2018empirical}, also known as the feedforward WaveNet~\cite{rethage2018wavenet}, that achieves a significant receptive field whilst operating on audio waveforms with the use of rapidly growing dilation factors~\cite{steinmetz2021efficient}. 
With appropriate data augmentation, our proposed model, WaveBeat, achieves comparable results to a previous deep learning approach~\cite{bock2020deconstruct}, even outperforming this approach on some datasets. 
This indicates not only are end-to-end models feasible for this task, they may provide a pathway for improved performance.
However, while these results are promising, they indicate our model struggles to generalize to unseen data distributions, lagging behind spectrogram-based approaches. 

\section{Proposed Model}\label{sec:model}


Estimating the location of the downbeat often requires significant context, potentially upwards of 30 seconds~\cite{fuentes2019music}. 
Constructing a model with a context window or receptive field of this size for waveforms requires attending to over 1 million timesteps, which imposes significant compute and memory cost.
The challenge in constructing an efficient end-to-end network has likely been a dominating factor in the use of sub-sampled spectral features in previous deep learning beat tracking approaches.
To address this challenge, our proposed model incorporates two core design elements: convolutions with rapidly growing dilation patterns and carefully designed subsampling. 

\subsection{Architecture}

The WaveBeat architecture is based on the TCN (or feedfordward WaveNet) design with a number of modifications.
The block diagram in Figure~\ref{fig:arch} demonstrates the overall structure at three different levels. 
Starting from the lowest level, on the right, each block is composed of residual 1-dimensional convolutional layers that incorporate batch normalization followed by a PReLU activation~\cite{he2015delving}.
The center of Figure~\ref{fig:arch} shows a TCN stack, which is composed of four blocks, after which the dilation pattern is repeated, a common approach~\cite{rethage2018wavenet}.
The complete model is shown on the left, composed of two TCN stacks, a $1\times1$ convolution to downmix to two channels, and a sigmoid function to generate the beat and downbeat activations.

TCNs commonly employ a dilation pattern such that the dilation factor at each layer $l \in \{1,2,...,N\}$ within a stack is given by $d_l = 2^{l-1}$, a convention likely the result of the approach introduced in~\cite{yu2016dilated}, later popularized for audio with  WaveNet~\cite{oord2016wavenet}.
To achieve an even larger receptive field, we consider utilizing a dilation pattern that grows more rapidly. 
Inspired by recent work in audio effect modeling~\cite{steinmetz2021efficient}, 
we consider increasing the growth such that the dilation factor at each layer is given by $d_l = 8^{l-1}$. 

We also note that is not useful to produce a beat activation function at audio sample rates, as the beat annotations are likely only accurate within tens of milliseconds~\cite{dixon2007evaluation}. 
To address this, we downsample the signal through the depth of the network by employing strided convolutions, similar to the approach used in a TCN-based encoder applied to room acoustics analysis~\cite{steinmetz2021filtered}.
With 8 layers, each with stride 2, we downsample the signal by a factor of $2^8 = 256$, which, given an input sample rate of 22.05\,kHz produces an output signal with a sample rate of 86\,Hz, close to those of previous works, which tend to be around 100\,Hz~\cite{bock2016joint}. 



\subsection{Loss function}

It is common to treat the beat tracking problem as a binary classification task. 
This is achieved by constructing a target signal $y_n$ with a value of 1 at each timestep $n$ containing a beat, and 0 elsewhere.
Then the model is then trained with the binary cross-entropy
\begin{equation*}
    \mathcal{L}_{\text{BCE}}(\hat{y},y) = -\frac{1}{N} \sum_{n=1}^{N} ({\hat{y}_n \log(y_n) + (1-\hat{y}_n) \log (1-y_n)}),
\end{equation*}\label{eq:bce}where the output of the model $\hat{y}_n$ is an estimate of the likelihood of a beat at the timestep $n$,
with the total number of timesteps $N$.
This can lead to a class imbalance across the temporal dimension, since there are often more locations with no beat. 
In practice, we found this encourages the model to avoid detecting beats, since such a solution will minimize the loss due to the small number of beat activations. 

To address this, we adapt the mean false  error $\mathcal{L}_{\text{MFE}}$~\cite{wang2016training}, a metric introduced to handle such class imbalances. 
Based upon the binary cross-entropy, we compute the loss as a sum of two terms that relate to the false-positive error and false-negative error\begin{align*}
    \mathcal{L}_{\text{FPE}}(\hat{y},y) &= \frac{1}{|J|} \sum_{j \in J} \mathcal{L}_{\text{BCE}}(\hat{y_j},y_j) \\
    \mathcal{L}_{\text{FNE}}(\hat{y},y) &= \frac{1}{|K|} \sum_{k \in K} \mathcal{L}_{\text{BCE}}(\hat{y_k},y_k) \\
    \mathcal{L}_{\text{MFE}} &= \mathcal{L}_{\text{FPE}} + \mathcal{L}_{\text{FNE}},
\end{align*}where $J$ is the set of timesteps corresponding to negative examples (no beat), and $K$ corresponds to the positive examples (beat).
This loss function attempts to balance performance by computing the sum of the average error at locations where a beat should be present, as well as the average error where there should be no beat. 
This encourages the model to avoid only predicting the majority class, i.e. the absence of a beat. However, we find as the sample rate of the beat activation function is reduced, this becomes less of an issue. 

\setlength{\tabcolsep}{0.5em}
\renewcommand{\arraystretch}{0.9}
\begin{table*}[t]
    \vspace{-0.4cm}
    \centering
    \begin{tabular}{l l l c c c c c c } \toprule
         & & &  \multicolumn{3}{c}{\textbf{Beat}} & \multicolumn{3}{c}{\textbf{Downbeat}} \\  \cmidrule(lr){4-6} \cmidrule(lr){7-9}
        \textbf{Dataset} & \textbf{Size} &  \textbf{Model}  & F-measure & CMLt & AMLt  & F-measure & CMLt & AMLt \\\midrule
        
        \multirow{3}{*}{\emph{Ballroom}} & \multirow{3}{*}{5 h 57 m} & Spectral TCN \cite{bock2020deconstruct} & \textbf{0.962} & \textbf{0.947} & \textbf{0.961} & 0.916 & 0.913 &  \textbf{0.960}  \\
        & & WaveBeat (Peak) & 0.961 & 0.929 & 0.929 & 0.904 & 0.762 & 0.803   \\ 
        & & WaveBeat (DBN) & 0.925 & 0.829 & 0.937 & \textbf{0.953} & \textbf{0.916} & 0.941  \\ \arrayrulecolor{black!30}\midrule
    
        \multirow{3}{*}{\emph{Hainsworth}} & \multirow{3}{*}{3 h 19 m} & Spectral TCN \cite{bock2020deconstruct} & 0.902 & 0.848 & 0.930 &  0.722 & 0.696 & 0.872  \\
        & & WaveBeat (Peak) & 0.965 & 0.937 & 0.937 & 0.912 & 0.748 & 0.843    \\ 
        & & WaveBeat (DBN) & \textbf{0.973} & \textbf{0.976} & \textbf{0.976} & \textbf{0.954} & \textbf{0.886} & \textbf{0.970}   \\ \arrayrulecolor{black!30}\midrule
     
        \multirow{3}{*}{\emph{Beatles}} & \multirow{3}{*}{8 h 09 m} & Spectral TCN \cite{bock2020deconstruct} & - & - & - & \textbf{0.837} & \textbf{0.742} & \textbf{0.862}\\
        & & WaveBeat (Peak) & 0.887 & 0.733 & 0.790 & 0.689 & 0.327 & 0.585   \\ 
        & & WaveBeat (DBN) & \textbf{0.929} & \textbf{0.894} & \textbf{0.894} & 0.732 & 0.509 & 0.724   \\ \arrayrulecolor{black}\midrule
        
        \multirow{3}{*}{\emph{GTZAN}} & \multirow{3}{*}{8 h 20 m} & Spectral TCN \cite{bock2020deconstruct} & \textbf{0.885} & \textbf{0.813} & \textbf{0.931} & \textbf{0.672} & \textbf{0.640} & \textbf{0.832} \\
        & & WaveBeat (Peak) & 0.825 & 0.682 & 0.767 & 0.563 & 0.279 & 0.515  \\ 
        & & WaveBeat (DBN) & 0.828 & 0.719 & 0.860 & 0.598 & 0.503 & 0.764  \\ \arrayrulecolor{black!30}\midrule
        
        \multirow{3}{*}{\emph{SMC}} & \multirow{3}{*}{2 h 25 m} & Spectral TCN \cite{bock2020deconstruct} & \textbf{0.544} & \textbf{0.443} & \textbf{0.635} & - & - & - \\
        & & WaveBeat (Peak) & 0.403 & 0.163 & 0.255 & - & - & -   \\ 
        & & WaveBeat (DBN) & 0.418 & 0.280 & 0.419 & - & - & - \\ 
        \arrayrulecolor{black} \bottomrule 
    \end{tabular}
    \caption{Beat and downbeat tracking results on the held-out test sets. No examples from the \emph{GTZAN} and \emph{SMC} datasets were seen during training.} \vspace{-0.3cm}
    \label{tab:results}
\end{table*}

\section{Experiments}\label{sec:experiments}

\subsection{Datasets}

In order to investigate the performance of the proposed model across a number of styles and audio sources, we consider six popular beat tracking datasets: \emph{Beatles}~\cite{davies2009evaluation}, \emph{Hainsworth}~\cite{hainsworth2004particle}, \emph{Ballroom}~\cite{gouyon2006experimental}~\cite{krebs2013rhythmic}, 
\emph{RWC Popular}~\cite{goto2002rwc},
\emph{SMC}~\cite{holzapfel2012selective}, 
and \emph{GTZAN}~\cite{tzanetakis2002musical, marchand2015gtzan}. 
Similar to previous works, we train using four datasets (\emph{Beatles}, \emph{Hainsworth}, \emph{Ballroom}, and 
\emph{RWC Popular}), and evaluate using two datasets that were not seen during training (\emph{SMC} and \emph{GTZAN}). 
All audio is resampled to $f_s = 22.05$\,kHz.

\subsection{Training}

We train WaveBeat where each convolutional layer utilizes kernels of size 15 and a stride of 2. The number of convolutional channels begins at 32 and then increases by 32 at each layer.
Combined with the rapidly growing dilation factors, this enables a receptive field of over 1 million timesteps, $\approx 47\,s$, using only 8 convolutional layers.
This is comparable to previous spectrogram-based beat tracking models that achieve a receptive field of around one minute~\cite{bock2019multi}. 
However, WaveBeat has a total of 2.9\,M trainable parameters, which is an order of magnitude more than common spectrogram-based models.

We utilize Adam with an initial learning rate of $1e^{-3}$, decreasing the learning rate by a factor of 10 after the beat and downbeat F-measure has not improved on the validation set for 10 epochs.
To stabilize training we apply gradient clipping when the norm of the gradients exceeds 4.
All models are trained with a batch size of 16 with inputs of $2^{21} = 2097152$\,samples ($\approx$1.6\,min at 22.05\,kHz) for a total of 100 epochs.
In order to balance the influence of the datasets while training, we define a single epoch to constitute 1000 random excerpts with replacement from each dataset. 
Additionally, we use automatic mixed precision to decrease training time and memory consumption. 
To facilitate reproducibility, we have made the code for these experiments  available online\footnote{\url{https://github.com/csteinmetz1/wavebeat}}.

\subsection{Data augmentation}

End-to-end approaches are more expressive than their counterparts that rely upon spectral features, thus they often require significantly more training data~\cite{pons2017end}. 
Due to the limited music recordings with beat and downbeat annotations, we found data augmentation critical in curbing overfitting.
We employ a set of fairly common data augmentations, 
each of which has an associated probability $p$ of being applied to each training example during a training epoch. 
This includes the applications of highpass and lowpass filters with random cutoff frequencies ($p=0.25$), random pitch shifting between -8 and 8 semitones ($p=0.5$), additive white noise ($p=0.05$), applying a $\tanh$ nonlinearity ($p=0.2$), shifting the beat locations forward or back by a random amount between $\pm$ 70\,ms ($p=0.3$), dropping a contiguous block of audio frames and beats of no more than 10\% of the input ($p=0.05$), as well as a random phase inversion ($p=0.5$).

\subsection{Post-processing}

Existing beat tracking systems generally utilize a post-processing stage which inspects the beat activation functions in order to select beat locations, commonly a dynamic Bayesian network (DBN)~\cite{bock2016joint}.
Ideally, an end-to-end model would be able to forgo such post-processing.
We analyze the beat and downbeat activation functions from WaveBeat using first simple peak picking, selecting peaks with an amplitude greater than 0.5. 
We then compare against beat activations produced by further post-processing with the pre-trained DBN in the \texttt{madmom} library~\cite{bock2016madmom} in order to examine the performance improvement.



\section{Evaluation}


We split the four training datasets into train/val/test sets (80\%/10\%/10\%). We utilize the standard distance threshold of $\pm$70\,ms and report the F-measure, CMLt, and AMLt metrics~\cite{davies2009evaluation} for both beat and downbeat tracking on the test sets in Table \ref{tab:results}.
We also show the reported scores for a recent spectrogram-based TCN model~\cite{bock2020deconstruct} as a point of comparison.
However, it should be noted that their scores are the result of an 8-fold cross validation, whereas our scores are computed using a single dataset split as described above.
Also, their model was trained with an additional three datasets (\emph{SMC}~\cite{holzapfel2012selective}, \emph{HJDB}~\cite{hockman2012one}, and \emph{Simac}~\cite{gouyon2005computational}), amounting to an additional 9 hours of training data, yet we employed more extensive data augmentation.
In the bottom of Table \ref{tab:results} we report results on the left-out datasets (\emph{GTZAN} and \emph{SMC}), in order to test the generalization capability.

Our results demonstrate that an end-to-end approach operating directly on waveforms can in fact achieve results on-par with current state-of-the-art approaches that employ carefully engineered input features. 
On the \emph{Ballroom} dataset we find that WaveBeat achieves comparable results on the beat tracking task, but achieves an improvement of 4\% in downbeat tracking. 
Similarly, WaveBeat achieves an improvement of over 7\% and 23\% on the \emph{Hainsworth} dataset for beat and downbeat tracking, respectively. 
While WaveBeat produces strong results on beat tracking on the \emph{Beatles} dataset, its performance is somewhat worse on downbeat tracking. 

However, these results indicate that WaveBeat falls behind the previous approach when generalizing to out-of-distribution examples and achieves comparable, yet lower results on the \emph{GTZAN} dataset on both beat and downbeat tracking. 
On the \emph{SMC} dataset, which contains a number of challenging pieces, WaveBeat performs clearly worse than the previous approach on beat tracking.
Results on the downbeat tracking task are omitted for this dataset due to the absence of downbeat annotations. 

With respect to the post-processing, we find that while the pre-trained DBN brings about a small improvement in the F-measure, 
our end-to-end model achieves comparable performance using simple peak picking.
This contrasts with previous approaches, which often report up to 15\% improvement with such post-processing~\cite{bock2016joint}.
Surprisingly, the results in Table~\ref{tab:results} appear to indicate that applying the DBN actually harmed performance in the case of beat tracking on the \textit{Ballroom} dataset.




\section{Conclusion}\label{sec:conclusion}

We demonstrated the ability of an end-to-end model to learn directly from waveforms on the joint beat and downbeat tracking task.
With an architecture designed to efficiently achieve a large receptive field,
we find that our model is able to achieve performance on-par with state-of-the-art methods for beat and downbeat tracking on some common datasets. 
We additionally investigate the requirement for specialized post-processing in the task of locating beat activations and find that our model performs well without such post-processing using very simple peak picking.
While these results are promising, indicating that end-to-end waveform based approaches can bring improvement over existing spectrogram-based methods, additional work is needed to improve the generalization ability of these approaches.
Future work involves the addition of a more rigorous data augmentation strategy, along with the application of self-supervised learning to leverage large music corpora without annotations. 

\section{Acknowledgement}
This work is supported by the EPSRC UKRI Centre for Doctoral Training in Artificial Intelligence and Music (EP/S022694/1).

\bibliographystyle{jaes}

\bibliography{refs}

\end{document}